\documentclass[a4paper,UKenglish,cleveref, autoref, thm-restate]{lipics-v2021}

\hideLIPIcs  

\usepackage{algorithm}
\usepackage{algpseudocode}
\usepackage{cleveref}
\title{Subexponential Algorithms for Continuous Euclidean \(k\)-Means in Subpolylogarithmic Dimension} 

\author{Junqi Tan}{Institut für Informatik, Freie Universität Berlin, Germany}{tanjunqi@zedat.fu-berlin.de}{}{}

\authorrunning{J. Tan}
\Copyright{Junqi Tan}


\category{} 

\relatedversion{} 

\nolinenumbers 

\EventEditors{John Q. Open and Joan R. Access}
\EventNoEds{2}
\EventLongTitle{42nd Conference on Very Important Topics (CVIT 2016)}
\EventShortTitle{CVIT 2016}
\EventAcronym{CVIT}
\EventYear{2016}
\EventDate{December 24--27, 2016}
\EventLocation{Little Whinging, United Kingdom}
\EventLogo{}
\SeriesVolume{42}
\ArticleNo{23}
\AtBeginDocument{%
  \hypersetup{pdfsubject={Exact algorithms for continuous Euclidean k-means}}%
}

\begin{document}

\maketitle

\begin{abstract}
Given $n$ points in $\mathbb{Q}^d$ and an integer $k$, the continuous
Euclidean $k$-means problem asks for $k$ centers in $\mathbb{R}^d$
that minimize the sum of squared distances from the input points to
their nearest centers.
In one dimension, the problem can be solved in polynomial time by dynamic programming. In higher dimensions,
the algorithm of Inaba, Katoh, and Imai [SoCG 1994] runs in polynomial
time when both $k$ and $d$ are fixed. Mahajan, Nimbhorkar, and Varadarajan [TCS 2012] showed that the problem is NP-hard even in the plane when $k$ is part of the input, answering a question of Dasgupta [2008].

In this paper, we give a deterministic exact algorithm for continuous Euclidean $k$-means. For arbitrary $k$ and $d=(\log n)^{o(1)}$, the algorithm runs in $O^*(2^{o(n)})$ time and uses polynomial space.

Aman, Karthik C.~S., and Punna [ESA 2024] asked
whether the $2^n$ time bound for Euclidean $k$-means can be improved
in high dimensions. Our result gives an affirmative answer to the corresponding question in subpolylogarithmic dimension.
\end{abstract}

\section{Introduction}
Given a set $P$ of $n$ points in $\mathbb{Q}^d$ and an integer $1\le k\le n$, the continuous Euclidean $k$-means problem asks for $k$ centers that minimize the sum of squared distances from the input points to their nearest centers. The centers may lie anywhere in $\mathbb{R}^d$. For a fixed cluster, the sum of squared distances to its center is minimized at the mean of its points. Equivalently, the problem asks for a partition of $P$ into $k$ nonempty clusters that minimizes the sum of squared distances from each point to the mean of its cluster.

The original motivation of Euclidean $k$-means comes from signal processing. In 1957, Lloyd ~\cite{DBLP:journals/tit/Lloyd82} studied how to represent signal amplitudes using only $k$ distinct values while minimizing the mean squared error. As computers made it practical to analyze larger data sets, Euclidean $k$-means found applications in tasks such as selecting color palettes
for digital images~\cite{DBLP:journals/ivc/Celebi11}, grouping genes with similar
expression profiles~\cite{tavazoie1999systematic}, constructing
codebooks for approximate nearest-neighbor search~\cite{DBLP:journals/pami/JegouDS11} and others. In modern AI systems, $k$-means arises as a subproblem in learning
visual features from unlabeled images~\cite{DBLP:conf/eccv/CaronBJD18}, generating
prediction targets for self-supervised speech models~\cite{DBLP:journals/taslp/HsuBTLSM21},
and compressing neural networks by quantizing their
weights~\cite{han2015deep}.

Euclidean $k$-means has been studied extensively in both the exact and
approximate settings. In one dimension, there is an optimal partition
into $k$ consecutive blocks of the sorted points, and such a partition
can be found in $O(kn)$ time after sorting~\cite{DBLP:journals/jal/Wu91}.
When both $k$ and $d$ are fixed, Inaba, Katoh, and
Imai~\cite{DBLP:conf/compgeom/InabaKI94} gave an exact algorithm that enumerates
Voronoi partitions using $O(n^{kd+1})$ arithmetic operations.
In the plane, Mahajan, Nimbhorkar, and
Varadarajan~\cite{DBLP:journals/tcs/MahajanNV12} showed that the
problem is NP-hard when $k$ is part of the input, answering a question
of Dasgupta~\cite{dasgupta2008hardness}. When $d$ is part of the input, the
problem is NP-hard even for $k=2$~\cite{dasgupta2008hardness}.
For every fixed $\varepsilon>0$, the problem admits a polynomial-time
$(1+\varepsilon)$-approximation when $k$ is
fixed~\cite{kumar2004simple} or when $d$ is
fixed~\cite{DBLP:journals/siamcomp/Cohen-AddadKM19}. When both $d$ and $k$ are part of the input, computing a $(1+\varepsilon_0)$-approximation is NP-hard for some constant $\varepsilon_0 > 0$ ~\cite{DBLP:conf/compgeom/AwasthiCKS15}.

A straightforward exact algorithm enumerates all $k^n$ assignments
of the input points to $k$ labeled clusters. Alternatively, a subset
dynamic program computes, for each $S\subseteq P$ and $1\le j\le k$,
the minimum cost of partitioning $S$ into $j$ nonempty clusters.
It considers each nonempty subset $T\subseteq S$ as one cluster
and adds its cost to the optimal cost of partitioning $S\setminus T$
into $j-1$ clusters. Since there are $3^n$ pairs $(S,T)$ with
$T\subseteq S\subseteq P$, the algorithm runs in $O^*(3^n)$ time
and uses $O^*(2^n)$ space in arbitrary dimension.

Aman, Karthik C.~S., and Punna~\cite{DBLP:conf/esa/AmanSP24} stated
that fast subset convolution yields a $2^n\operatorname{poly}(n,d)$-time
algorithm for Euclidean $k$-means. They asked whether this bound can
be improved in high dimensions. For $k=2$, they gave an $O^*(1.7297^n)$-time
exact algorithm in arbitrary dimension, assuming that all coordinates
are integers of absolute value at most $2^{o(n)}$. More recently, Bai, Fomin, and Xu~\cite{BaiFominXu2026} gave a
deterministic exact algorithm for planar Euclidean $k$-means.
On rational inputs, it runs in $n^{O(k/(\log k)^{2/3})}$ time up to
a polynomial factor in the input encoding length. This improves the
exponent of $n$ from linear to sublinear in $k$. When $k$ is linear in $n$, however, this bound does not guarantee
subexponential time in $n$. Can we obtain a subexponential-time
algorithm when the dimension is a constant and $k$ is unrestricted?
We show that the answer is yes for every fixed dimension and,
more generally, for $d=(\log n)^{o(1)}$. To our knowledge, no previous exact algorithm for continuous Euclidean \(k\)-means achieves \(O^*(2^{o(n)})\) worst-case running time for unrestricted \(k\) throughout this
range of dimensions.

In this paper we show the following main result. 

\begin{theorem}\label{thm1}
Let $P\subseteq\mathbb{Q}^d$ be a set of $n$ points, and let
$1\le k\le n$ be an integer. If $d=(\log n)^{o(1)}$, then
there is a deterministic algorithm that finds an optimal
continuous Euclidean $k$-means partition of $P$ in
$O^*(2^{o(n)})$ time and uses polynomial space.
\end{theorem} 

The theorem allows arbitrary rational coordinates given in binary.
Their numerators and denominators may have arbitrary bit length.
The $O^*(\cdot)$ notation suppresses a factor polynomial in the
total binary encoding length of the input, and the space bound
is polynomial in the same quantity.
\subparagraph{Overview of our approach.}

Fix an optimal partition of the input points, which is not known to the algorithm. We consider $K$ disjoint intervals of cluster sizes, where $K$ grows slowly with $n$. At least one interval contains clusters with at most $n/K$ points in total. We enumerate their union $Z$ and solve the instance on $Z$ using a brute-force algorithm. For the correct choice of $Z$, every remaining cluster has size at most $L$ or greater than $U$, for suitable values $L<U$. We call these clusters \emph{small} and \emph{large}, respectively.

For the small clusters, our main observation is that the centers of all but possibly one small cluster are sufficiently far apart. Otherwise, we could split one small cluster and merge two nearby small clusters to obtain a solution with smaller cost. Using this property, we construct a shifted grid and remove a small union $B$ of complete small clusters. The remaining points can then be divided into blocks of bounded size such that every remaining small cluster lies entirely within one block. This allows us to handle the small clusters separately within each block.

The difficulty is that large clusters may contain points from several blocks. We assign a common label to the points belonging to the same large cluster. If the centers of the large clusters were known, we could choose a minimum-cost assignment independently in each block. To avoid guessing these centers, we introduce additional scalar parameters that make the scores of the assignments affine. We then use hyperplanes to divide the parameter space into regions and generate one parameter point from each region. We show that these points are sufficient to find an optimal combination of assignments.

Finally, the gap between $L$ and $U$ allows us to bound both the size of each block and the number of large clusters. By choosing $K$ appropriately, we show that all the required choices can be enumerated in $O^*(2^{o(n)})$ time and polynomial space when $d=(\log n)^{o(1)}$.

\section{Preliminaries}

Let $P=\{p_1,\ldots,p_n\}$ be a set of $n$ points in $\mathbb{Q}^d$, where $d=(\log n)^{o(1)}$. In the continuous Euclidean $k$-means problem, the goal is to find a
partition of $P$ into $k$ nonempty clusters that minimizes the sum of squared distances from each point to the center of its cluster.

For a nonempty subset $S\subseteq P$, let $\mu(S)$ denote the center of $S$, where $\mu(S)=\frac{1}{|S|}\sum_{p\in S}p$. We define the cost of $S$ by $\operatorname{cost}(S)$ as the sum of squared distances from each point of $S$ to $\mu(S)$.
\[
\operatorname{cost}(S)
=\sum_{p\in S}\|p-\mu(S)\|^2.
\]
Equivalently,
\begin{equation}
\operatorname{cost}(S)
=\sum_{p\in S}\|p\|^2
-\frac{1}{|S|}
 \left\|\sum_{p\in S}p\right\|^2.
\label{eq:cluster-cost}
\end{equation}

For a partition $\mathcal{P}$ of $P$, we define the cost of $\mathcal{P}$ as 
\[
\operatorname{cost}(\mathcal{P})
=\sum_{S\in\mathcal{P}}\operatorname{cost}(S).
\]

The following lemma shows how the cost changes when two clusters
are merged.

\begin{lemma}\label{merge-cost}
For any two disjoint nonempty subsets $A,B\subseteq P$,
\[
\operatorname{cost}(A\cup B)
=
\operatorname{cost}(A)+\operatorname{cost}(B)
+\frac{|A||B|}{|A|+|B|}
 \|\mu(A)-\mu(B)\|^2.
\]
\end{lemma}

\begin{proof}
Let $a=|A|$ and $b=|B|$.
By Equation~\eqref{eq:cluster-cost},
\begin{align*}
&\operatorname{cost}(A\cup B)
-\operatorname{cost}(A)-\operatorname{cost}(B)\\
&\quad=
a\|\mu(A)\|^2+b\|\mu(B)\|^2
-\frac{\|a\mu(A)+b\mu(B)\|^2}{a+b}\\
&\quad=
\frac{ab}{a+b}\|\mu(A)-\mu(B)\|^2.
\end{align*}
\end{proof}

We will also use a simple exact algorithm for solving
small instances.

\begin{lemma}[Brute-Force]
\label{lem:baseline}
Continuous Euclidean $k$-means can be solved in $O^*(2^{3n+O(\log^2(n))})$ time and polynomial space.
\end{lemma}

\begin{proof}
Consider an optimal partition of $P$. If there exists a cluster larger than $2n/3$,
we guess it and recursively solve the continuous Euclidean $k-1$-means on the remaining instance, which
contains fewer than $n/3$ points. Otherwise, there exists a union $X$ of some clusters such that $\frac{n}{3}\le |X|\le\frac{2n}{3}$.

We consider both types of decompositions. For the second type, we guess a subset $X$ such that $\frac{n}{3}\le |X| \le\frac{2n}{3}$. For each $X$, we assign the $k$ clusters between $X$ and $P\setminus X$, and
recursively solve both subinstances. The two recursive calls are performed sequentially. Finally, we choose the optimal one among all the resulting partitions.

Since every candidate is a valid partition and at least one
enumerated decomposition is compatible with the chosen
optimal partition, the algorithm returns an optimal solution.

For the running time, consider a recursive call on $m$
points. We enumerate at most $2^m$ subsets and $m$ assignments of clusters. Each candidate requires at most
two recursive calls on at most $2m/3$ points. 
Let $n_0=n$, and let \(n_i\) be the maximum size of any subinstance at depth \(i\). Since
$n_{i+1}\le 2n_i/3$ and the recursion depth is at most $h=O(\log n)$, we have
\[
\sum_{i=0}^{h-1}n_i
\le
n+\frac23n+\left(\frac23\right)^2n+\cdots
\le 3n.
\]

Since $n_i\le n$, we get
\[\prod_{i=0}^{h-1}n_i\le n^{O(\log n)}.
\]

Thus, the running time is bounded by
\[
\prod_{i=0}^{h-1} (2\cdot 2^{n_i}\cdot n_i)
\le
2^{O(\log n)} \cdot 2^{3n}\cdot n^{O(\log n)}
\le 2^{3n+O(\log^2(n))}.
\]

Since all subsets are generated one at a time and each call stores only polynomial-size data, the algorithm uses polynomial
space.
\end{proof}

Throughout the paper, $d=(\log n)^{o(1)}$ and the $O^*$ notation suppresses
factors polynomial in the total input encoding length.

\section{The Algorithm}
Let $C_1,\ldots,C_k$ be an optimal partition of $P$ with
$|C_1|\le\cdots\le|C_k|$.
Our algorithm first guesses a union $Z$ of all clusters whose sizes lie in a suitable interval, for example, $Z=C_i\cup C_{i+1}\cdots\cup C_{i+j}$. 
We solve the resulting instance on $Z$ by the brute-force algorithm from \cref{lem:baseline}. 

After removing $Z$,  we use a shifted grid to divide the remaining points
into small blocks. We guess a union of some small clusters and solve it using the same brute-force algorithm. We show that some choice of the grid and this union
leaves every remaining small cluster entirely within one block.
For each block, we consider the possible assignments of points to small and large clusters. We solve the
small clusters by the same brute-force algorithm. Large clusters may contain points from several blocks. Finally, we use affine comparisons to combine the local choices and obtain an optimal partition.

\subsection{Creating a Size Gap}
\label{sec:size-gap}

We first guess a union of a subset of clusters such that the remaining clusters are either small or large. The case $n=1$ is immediate. For $n\ge 2$, let
\[
K=\left\lfloor\frac{\log_2\log_2 n}{2\log_2(d+2)}\right\rfloor.
\]
If $K<2$, we solve the instance using Lemma~\ref{lem:baseline}.
Otherwise, we iteratively generate $K$ disjoint intervals for the cluster sizes. Initially setting $U=n$, we define the interval as $(L,U]$ where $L=\lfloor U^{1/(d+2)}\rfloor$. We then update $U = L$ and repeat the process $K-1$ times. By this construction, each resulting interval satisfies $L^{d+2}\le U$. Fix an optimal partition of $P$. By the pigeonhole principle, at least one interval contains clusters of this partition whose total size is at most $n/K$.

Our algorithm enumerates all $K$ intervals and all subsets $Z\subseteq P$ with $|Z|\le \lfloor n/K\rfloor$. For each $Z$, we guess its number of clusters $t$ and solve the
resulting instance using Lemma~\ref{lem:baseline}. Note that for each call to the brute-force algorithm on a point
set $Y$, we discard a cluster count if $Y$ cannot be
partitioned into that many nonempty clusters.
For an empty set, the only feasible count is zero,
with cost zero.
For a nonempty set $Y$, the feasible counts are $1,\ldots,|Y|$. We then generate candidate partitions of $P\setminus Z$ using at most $k-t$ clusters, as described in the following subsections. For a correct guess, $Z$ is exactly the union of the clusters of the fixed optimal partition whose sizes lie in $(L,U]$. The remaining clusters form an optimal partition of $P\setminus Z$, with each cluster having size at most $L$ or greater than $U$. We call them small and large clusters, respectively. In the following subsections, we show how to handle these small clusters and large clusters.

\subsection{Handling Small Clusters}
\label{sec:small-clusters}

Consider a correct guess of $(L,U]$, $Z$, and $t$
from the previous subsection. Let $X=P\setminus Z$ and $m=|X|$. The
remaining clusters form an optimal partition of $X$, and each cluster has size at most $L$ or greater than $U$. In this section, we focus on how to handle the small clusters.

Assume that input points are distinct and have integer coordinate (the full details of the preprocessing for arbitrary rational inputs are provided in the appendix). 
First suppose that some small cluster contains at least two points.
Let $A$ be a small cluster of maximum radius, and let $R$ denote its
radius. Since the input points are distinct, $R>0$. We show that the
centers of any two small clusters other than $A$ are sufficiently far
apart.
\begin{lemma}
\label{lem:spacing}
For any two distinct small clusters $C$ and $D$
other than $A$,
\[
\|\mu(C)-\mu(D)\|\ge R\sqrt{2/L}.
\]
\end{lemma}

\begin{proof}
Choose a point $x\in A$ with $\Vert{}x-\mu(A)\Vert{}=R$. Consider modifying the partition by splitting $A$ into $\{x\}$ and $A\setminus\{x\}$ while merging $C$ and $D$. Splitting $A$ decreases the cost by
\[
\frac{\vert{}A\vert{}}{\vert{}A\vert{}-1}R^2\ge R^2
\]

Conversely, according to \cref{merge-cost}, merging $C$ and $D$ into $C\cup D$ increases the cost by
\[
\frac{\vert{}C\vert{}\vert{}D\vert{}}{\vert{}C\vert{}+\vert{}D\vert{}} \Vert{}\mu(C)-\mu(D)\Vert{}^2 \le \frac{L}{2}\Vert{}\mu(C)-\mu(D)\Vert{}^2.
\]

Since this modification preserves the number of clusters, the optimality of the partition implies
\[
\frac{L}{2}\Vert{}\mu(C)-\mu(D)\Vert{}^2\ge R^2
\]

This proves \cref{lem:spacing}
\end{proof}

We use an axis-parallel grid to divide the space into equal-sized $d$-dimensional cubes, called \emph{cells}. We show that, by choosing the grid appropriately, only a small number of points belong to small clusters crossing cell boundaries, and all other small clusters lie entirely within individual cells.

\subparagraph{Dividing the points by a grid.}
Let $a=\left\lfloor L^{1/[2(d+1)]}\right\rfloor$ and $s=L^{\lceil(d+3)/2\rceil}$.
We use an axis-parallel grid whose cells have side length
$H=a\rho$, where $R\le\rho\le2R$.
The parameter $a$ controls how many points must be removed,
and $s$ bounds the size of each final block.
We explain how to enumerate $\rho$ and the grid shift below.

\begin{lemma}
\label{lem:crossing}
There exists a shift of the grid such that the small clusters intersecting more than one cell contain at most $2dm/a$ points in total.
Every other small cluster lies entirely within one cell.
\end{lemma}
\begin{proof}
Shift the grid uniformly at random.
Each small cluster has a span of at most $2R$ in
each coordinate, so the probability that it intersects
more than one cell is at most
\[
\frac{2dR}{H}\le\frac{2d}{a}.
\]
Multiplying this probability by the size of each
small cluster and summing, the expected total number
of points in such clusters is at most $2dm/a$.
Hence, some shift satisfies the claimed bound.
\end{proof}

Fix such a shift. We set aside all small clusters crossing cell boundaries, so that every remaining small cluster lies entirely within one cell.
However, some cells may still contain too many points, including points belonging to large clusters. We next show how to handle these cells without
splitting any remaining small cluster.

Consider the small clusters entirely contained in a
cell of side length $H$. Their centers also lie in
this cell. By \cref{lem:spacing}, the centers of any two small
clusters other than $A$ are at distance at least
$\Delta=R\sqrt{2/L}$. Around each center, we place a cube of
side length $\Delta/\sqrt d$.
These cubes have disjoint interiors and lie within
a larger cube of side length $H+\Delta/\sqrt d$.
Hence, the number of such clusters, excluding $A$,
is at most
\[
\left(1+\frac{\sqrt d H}{\Delta}\right)^d
\le \left(1+a\sqrt{2dL}\right)^d,
\]
where the inequality follows from $H\le 2aR$.

Including $A$, each cell contains at most
\[
1+\left(1+a\sqrt{2dL}\right)^d
\le \varepsilon_d a^d L^{d/2}
\]
complete small clusters, where $\varepsilon_d=1+\left\lceil(1+\sqrt{2d})^d\right\rceil$.

We call a cell heavy if it contains more than $s$ points. Since there are $m$ points in total, there are fewer than $m/s$ heavy cells. We set aside all remaining small clusters contained
in heavy cells. By the bound above, the total number
of points set aside in this step is at most
\[
\frac{m}{s}\varepsilon_d a^d L^{1+d/2}
\le \frac{\varepsilon_d m}{a},
\]
where the inequality follows from
$a^{d+1}\le\sqrt{L}$ and
$s\ge L^{(d+3)/2}$.

Let $B$ be the union of all small clusters set aside
so far, including those crossing cell boundaries.
By \cref{lem:crossing}, we have
\[
|B|\le \frac{(2d+\varepsilon_d)m}{a}.
\]

After removing $B$, every cell either contains at
most $s$ points or contains only points belonging
to large clusters. In the latter case, we can
divide its points arbitrarily into sets of size
at most $s$.

We now have a collection of sets, each containing
at most $s$ points. We combine these sets using a greedy procedure that adds the next set
whenever it fits, so that each resulting set contains at most $s$ points.
We call the resulting sets \emph{blocks} and
denote them by $X_1,\ldots,X_f$. Thus, we obtain
\[
f\le \frac{2(m-|B|)}{s}+1.
\]
Every small cluster not contained in $B$ lies
entirely within one block.

The construction above depends on the unknown optimal
partition. We now show how to enumerate the grid and
the set $B$ without knowing this partition.

\subparagraph{Enumerating the grid and the removed points.} Recall that all input points have distinct integer
coordinates. If $R>0$, then $R\ge 1/2$.
We enumerate powers of two
$\rho=1,2,4,\ldots$ up to a value exceeding the
maximum possible radius. Thus, for some choice,
$R\le\rho\le 2R$.

For each $\rho$, we set $H=a\rho$ and enumerate
the possible shifts of the grid. Along each coordinate,
the cells containing the input points change only
when the shift passes a coordinate of an input point
modulo $H$. Hence, it suffices to consider at most
$m+1$ representative shifts in each coordinate,
giving at most $(m+1)^d$ shifts in total.

For each grid, we enumerate all subsets
$B\subseteq X$ with
\[
|B|\le
\min\left\{m,
\left\lfloor\frac{(2d+\varepsilon_d)n}{a}\right\rfloor
\right\}.
\]
We then construct the blocks $X_1,\ldots,X_f$
as described above.

By the preceding arguments, at least one choice
of the grid and $B$ leaves every small cluster
outside $B$ entirely within one block.
The algorithm does not need to identify this choice.

If every small cluster is a singleton, or if there
are no small clusters, we also consider $B=\emptyset$
and divide the points directly into blocks of size
at most $s$.

\subparagraph{Summary.} We have now divided $X$ into $B$ and
$X_1,\ldots,X_f$.
We guess the number of clusters in $B$ and solve
the instance on $B$ using the brute-force algorithm
from Lemma~\ref{lem:baseline}.

Each block $X_i$ may contain both small and large
clusters. Every small cluster lies entirely within
one block, while a large cluster may have points
in several blocks. In the next subsection, we show
how to determine the clusters within each block
and combine the parts of large clusters across
different blocks.

\subsection{Handling Large Clusters}
\label{sec:large-clusters}
Recall that $X$ has been divided into $B$ and $X_1,\ldots,X_f$, where every small cluster outside $B$ lies entirely within one block, while large clusters may span several blocks. Since we assume that $t$ is the number of clusters assigned to $Z$, $X$ has $k-t$ clusters.

We first guess the number $q$ of large clusters.
Since each large cluster contains more than $U$
points, $q$ should satisfy
\[
q(U+1)+(k-t-q)\le m.
\]
Let $C_1, \dots, C_q$ denote these large clusters.

In the previous section, we guess the number of small clusters in $B$ and solve the instance on $B$ using the brute-force
algorithm from Lemma~\ref{lem:baseline}. Assume that the number of small clusters in $B$ is $t_B$. We then guess the number of small clusters in each block. Let these numbers be
$t_1,\ldots,t_f$, respectively, satisfying
\[
t_B+\sum_{i=1}^f t_i=k-t-q.
\]

For each block $X_i$, we consider all assignments of its points to labels $0,1,\ldots,q$. Label $0$ means that a point belongs to a small cluster, while label $j>0$ means that it belongs
to $C_j$. We call an assignment \emph{feasible} if the points labeled $0$ can be partitioned into exactly $t_i$
nonempty clusters.
Thus, if $t_i=0$, no point may receive label $0$;
otherwise, at least $t_i$ points must receive this
label.
A positive label may receive no points in a block. If some block admits no feasible assignment, we discard the current
guess of the cluster counts.

For a fixed assignment, we solve the points labeled $0$ into $t_i$ clusters using the same brute-force algorithm. For each positive label, we combine all points carrying this label across different blocks.

For fixed $B$ and $t_B$, the cost of the solution on $B$ is independent of the assignments within the blocks. We therefore consider only the cost on $X\setminus B$ and add the cost on $B$ when evaluating the complete solution.

We now compute the cost of the partition of $X\setminus B$
obtained by choosing one feasible assignment from
each block.

\subparagraph{Computing the cost on $X\setminus B$.}
Consider a fixed assignment of the points in a block
$X_i$. For each positive label $j$, let $h_{ij}$ be
the number of points carrying this label, and let
$u_{ij}$ be the sum of their coordinates.
Let $b_i$ be the minimum cost of partitioning the
label-$0$ points into $t_i$ clusters, plus the sum
of squared norms of all positively labeled points.
The former is computed using the brute-force algorithm
from Lemma~\ref{lem:baseline}.
We consider only feasible assignments.
If no point is labeled $0$, the cost of the small clusters is zero.

Now choose one feasible assignment for each block.
For each label $j$, we combine all points carrying
this label into one cluster.
This cluster contains $\sum_i h_{ij}$ points, whose
coordinate sum is $\sum_i u_{ij}$.
If it is nonempty, its center is therefore
\[
\frac{\sum_{i=1}^f u_{ij}}
     {\sum_{i=1}^f h_{ij}}.
\]
Thus, we do not need to specify its center when
processing an individual block.

By Equation~\eqref{eq:cluster-cost}, the cost of
the resulting partition of $X\setminus B$ is
\begin{equation}
\sum_{i=1}^f b_i
-
\sum_{\substack{1\le j\le q\\
                \sum_{i=1}^f h_{ij}>0}}
\frac{\left\|\sum_{i=1}^f u_{ij}\right\|^2}
     {\sum_{i=1}^f h_{ij}}.
\label{eq:combined-cost}
\end{equation}
The squared norms of the positively labeled points
are already included in the values $b_i$.

Note that we impose no size restrictions on the returned
clusters. The size gap is used only to ensure that
our enumeration includes a choice compatible with
the fixed optimal partition.

It remains to find a combination of assignments
that minimizes Equation~\eqref{eq:combined-cost}. Enumerating all combinations
may be too expensive. We next show how to avoid this
by using affine comparisons.

\subparagraph{Choosing an assignment for each block.}
Suppose that the centers of the large clusters are
fixed at $z_1,\ldots,z_q\in\mathbb{R}^d$.
For a feasible assignment in block $X_i$, its
contribution to the cost on $X\setminus B$ is
\[
b_i+\sum_{j=1}^q
\left(
h_{ij}\|z_j\|^2-2\langle z_j,u_{ij}\rangle
\right).
\]
This follows by expanding the squared distances
to the fixed centers.
The total cost on $X\setminus B$ is the sum of
these contributions over all blocks.
Thus, for these fixed centers, we can choose a minimum-cost
assignment in each block independently.

For each point $x$ of $X_i$ assigned to label $j$, we have
\[
\|x-z_j\|^2
=
\|x\|^2-2\langle x,z_j\rangle+\|z_j\|^2.
\]
Thus, each such point contributes a term $\|z_j\|^2$.
Since there are $h_{ij}$ such points in $X_i$, these
terms sum to $h_{ij}\|z_j\|^2$.

We want to compare assignments within each block using
hyperplanes. To achieve this, we next define an affine score for each assignment. 
For two assignments at the same centers, we subtract
their costs. The constant terms and the inner-product terms
give an affine expression in the center coordinates.
However, the terms involving $\|z_j\|^2$ are not affine in $z_j$.
For example, suppose one assignment assigns two points of $X_i$ to label $j$, while another assigns three. Their cost difference then contains the term $-|z_j|^2$, which is not affine in $z_j$. 

We therefore introduce an additional scalar parameter
$w_j$ for each large-cluster label $j$ and replace
$\|z_j\|^2$ by $w_j$.
We treat $w_j$ and the
coordinates of $z_j$ as independent parameters.
The same pair $(z_j,w_j)$ is used for label $j$ in all
assignments and all blocks.
For each assignment in block $X_i$, we define the affine score for it as follows
\begin{equation}
g_i(z,w)
=
b_i+\sum_{j=1}^q
\left(
h_{ij}w_j-2\langle z_j,u_{ij}\rangle
\right).
\label{eq:local-score}
\end{equation}

In the above example, the two terms now become $2w_j$ and $3w_j$,
whose difference is $-w_j$ (a linear term).
Fix an assignment, the coefficients
$b_i$, $h_{ij}$, and $u_{ij}$ are constants. Each center $z_j$ contributes $d$ coordinate parameters,
and $w_j$ contributes one additional parameter.
Thus, there are $(d+1)q$ scalar parameters in total.
Since the score consists of a constant plus terms linear
in these parameters,
it is affine. We will explain later how these affine scores allow us to compare different assignments using hyperplanes.

Once the parameters $z_1,\ldots,z_q\in\mathbb{R}^d$
and $w_1,\ldots,w_q\in\mathbb{R}$ are fixed,
 we can choose a minimum-score assignment independently
in each block. For each block, we enumerate the assignments. For each feasible assignment,  we compute its score at these parameters.
We then choose an assignment with the smallest score among all feasible assignments.
If several assignments have the same minimum score,
we choose any of them.

Now we show that there exists a choice
of these parameters such that combining any minimum-score
assignment from each block gives a minimum-cost partition
of $X\setminus B$ among all partitions obtained from
feasible assignments.

\begin{lemma}
\label{lem:local-choices}
Fix the blocks $X_1,\ldots,X_f$, the number $q$ of
large-cluster labels, and the numbers $t_1,\ldots,t_f$
of small clusters in the blocks. Suppose that every block admits at least one feasible
assignment.
Then there exist parameters $z_1,\ldots,z_q\in\mathbb{R}^d$
and $w_1,\ldots,w_q\in\mathbb{R}$ with the following property.
Choose any minimum-score assignment in each block
and combine these assignments into a partition of
$X\setminus B$.
The resulting partition has minimum cost among all
partitions obtained by combining feasible assignments.
\end{lemma}

\begin{proof}
Fix a combination of feasible assignments with
minimum total cost.
We label the large clusters by $1,\ldots,q$,
adding empty clusters if there are fewer than $q$.
For each $j\in\{1,\ldots,q\}$, if the $j$-th cluster
is nonempty, let $z_j$ be the average of the coordinates
of its points. Otherwise, let $z_j=0\in\mathbb{R}^d$.
For every $j$, let $w_j=\|z_j\|^2$.

With these parameters, the sum of the scores of the
assignments in the original combination equals its
total cost.
Suppose that some block has another feasible assignment
with a strictly smaller score.
Replacing the original assignment in this block with
the new one strictly decreases the total cost at the
fixed centers $z_1,\ldots,z_q$.
Since recomputing the centers of the large clusters can not increase
this cost, we obtain a feasible combination with
strictly smaller total cost, contradicting the optimality
of the original combination.
Hence, every original assignment has minimum score
in its block.

Now we choose a minimum-score assignment in each block (it may be different from the original assignment). 
The sum of the scores is unchanged. Thus, the cost at the
fixed centers $z_1,\ldots,z_q$ equals the cost of the
original combination.
Recomputing the centers of large clusters
can not increase this cost.
Thus, the cost of the new combination is not larger than the cost of the original combination. Therefore, the new combination is optimal.
\end{proof}

Since these parameters have infinitely possible values, we cannot enumerate them. Instead, we use the hyperplanes defined by equal scores of assignments within the same block. These hyperplanes divide the parameter space into open regions. Within each region, the order of the scores remains unchanged, so each block has the same minimum-score assignments. We will show that it is sufficient to generate one parameter point from each region.

\subparagraph{Generating parameters.}
Let $r=(d+1)q$. 
Let $\theta=(z_1,\ldots,z_q,w_1,\ldots,w_q)$
denote the vector containing all coordinates of the centers of the large clusters
and the additional scalar parameters.
Thus, $\theta\in\mathbb{R}^r$. Consider two feasible assignments $\alpha$ and $\beta$ in the same block $X_i$.
Since their scores are affine in the parameters $\theta$, their difference is also affine.
If this difference is not constant, the values of the parameters for which the two scores are equal form a hyperplane. For example, suppose $d=q=1$. Thus, $\theta=(z_1,w_1)$. Suppose $\alpha$ and $\beta$ in block $X_i$ have the following scores:
\[
g_i^\alpha(\theta)=1+2w_1-2z_1,
\qquad
g_i^\beta(\theta)=\frac92+w_1-4z_1.
\]
Their scores are equal exactly when
\[
w_1+2z_1-\frac72=0,
\]
which defines a line in the two-dimensional parameter space.
On one side of this line, $\alpha$ has a smaller score; on the other side, $\beta$ has a smaller score.
For general $d,q$, such equalities define hyperplanes in $\mathbb{R}^r$.
Note that the hyperplane is not needed when the score difference is constant. Thus, each hyperplane can be represented by an equation
\[
a\cdot\theta+c=0,
\]
where $a\in\mathbb{Q}^r$ and $c\in\mathbb{Q}$.

To generate the parameters, we introduce one additional coordinate $t$ and consider vectors $v=(y,t)\in\mathbb{R}^{r+1}$, where $t>0$ and $y\in\mathbb{R}^r$. We recover the original parameters by setting $\theta=y/t$.
For each hyperplane,
we introduce two row vectors $(a,c)$ and $-(a,c)$.
Since
\[
a\cdot\theta+c
=\frac{a\cdot y+ct}{t}.
\]
and \(t>0\), the signs of \(a\cdot\theta+c\) and \((a,c)\cdot v\) are the same. Thus, the two row vectors correspond to the two sides of the hyperplane.
We also include the row $(0,\ldots,0,1)$ to represent the condition $t\ge1$ used in the proof below.

We consider all the row vectors $(a,c)$ and $-(a,c)$
obtained from the hyperplanes, together with the
additional row $(0,\ldots,0,1)$.
We index these rows and enumerate all subsets
containing between $1$ and $r+1$ rows. For each subset, we first check whether the selected
rows are linearly independent.
If they are not, we skip this subset.
Otherwise, we construct a matrix $A$ by placing the
selected rows one below another. We then solve
\begin{equation}
(AA^\top)\lambda=\mathbf{1},
\qquad
v=A^\top\lambda,
\label{eq:sample-system}
\end{equation}
where $\mathbf{1}$ is the all-ones vector with
as many entries as there are rows in $A$.
Since the rows of $A$ are linearly independent,
$AA^\top$ is invertible. Thus, $\lambda$ and $v$
are uniquely determined.

Let $v_{j}$ denote the $j$-th coordinate of $v$.
If $v_{r+1}\le0$, we skip the current subset.
Otherwise, we compute the parameter vector
\[
\theta=\frac{(v_1,\ldots,v_r)}{v_{r+1}}.
\]
If $\theta$ lies on any of the hyperplanes,
we discard it. Otherwise, we keep it as a sample.

We repeat this process for every enumerated subset.
If there are no hyperplanes, we simply choose $\theta=0$.

We next prove that this procedure generates at least
one parameter vector in every nonempty open region.

\begin{lemma}
\label{lem:comparison-samples}
Every nonempty open region defined by the hyperplanes
contains at least one generated sample.
\end{lemma}

\begin{proof}
Fix an arbitrary nonempty open region determined by the hyperplanes.
For each hyperplane
\[
a_i\cdot\theta+c_i=0,
\]
the expression $a_i\cdot\theta+c_i$ is either
strictly positive or strictly negative throughout
this region.
Let $\sigma_i\in\{+1,-1\}$ denote its sign. Thus, for every $i$ this region consists exactly of the points
$\theta\in\mathbb{R}^r$ satisfying
\[
\sigma_i(a_i\cdot\theta+c_i)>0
\]

We now show that every open region contains a point
that can be obtained from a solution to a system of
non-strict linear inequalities. Consider the following system of inequalities for $t \ge 1$:
\[
\sigma_i(a_i \cdot y + c_i t) \ge 1 \qquad \text{for all } i.
\]

We first show that these inequalities have a solution.
Take any point $\theta$ in the chosen region.
Since
\[
\sigma_i(a_i\cdot\theta+c_i)>0
\]
for every $i$, we can choose a sufficiently large
$\lambda\ge 1$ such that
$y=\lambda\theta$ and $t=\lambda$
satisfy all the inequalities.

Conversely, for any solution $(y,t)$, we have $t\ge 1$.
Dividing the inequalities by $t$ gives
\[
\sigma_i\left(a_i\cdot\frac{y}{t}+c_i\right)
\ge\frac{1}{t}>0.
\]
Thus, $\theta=y/t$ belongs to the chosen region.

Among all vectors $v=(y,t)$ satisfying these
inequalities, choose one with minimum Euclidean norm,
denoted by $v^*$. Since the inequalities have
a solution and their solution set is closed, such a vector exists .

We call a constraint tight if its left-hand side equals
$1$ at $v^*$.
We claim that $v^*$ is a linear combination of the
coefficient vectors of these tight constraints.

Suppose otherwise.
Then we can write $v^*=v_0+u$, where $v_0$ lies
in the span of the coefficient vectors of the
tight constraints, and $u\ne0$ is perpendicular
to all these vectors.
For sufficiently small $\varepsilon>0$, consider
\[
v'=v^*-\varepsilon u.
\]
Every tight constraint remains an equality because
$u$ is perpendicular to its coefficient vector.
All other constraints remain satisfied because
they hold with strict inequality at $v^*$.
Moreover, $\|v'\|<\|v^*\|$, contradicting the
choice of $v^*$.

Therefore, $v^*$ lies in the space spanned by
the coefficient vectors of the tight constraints.
Since $v^*\in\mathbb{R}^{r+1}$, we can choose
at most $r+1$ linearly independent tight constraints
whose coefficient vectors span this space.

Let $A$ be the matrix whose rows are these
linearly independent coefficient vectors.
Since all these constraints are tight at $v^*$,
we have
\[
Av^*=\mathbf{1}.
\]

Since $v^*$ lies in the row space of $A$,
we can write $v^*=A^\top\lambda$ for some vector
$\lambda$.
Substituting this into $Av^*=\mathbf{1}$ gives
\[
(AA^\top)\lambda=\mathbf{1}.
\]
The rows of $A$ are linearly independent,
so $AA^\top$ is invertible. Therefore,
\[
v^*=A^\top(AA^\top)^{-1}\mathbf{1}.
\]

Our algorithm enumerates every set of at most
$r+1$ linearly independent rows, so it will
consider this matrix $A$ and recover $v^*$.

Write $v^*=(y^*,t^*)$.
Since $v^*$ satisfies all the inequalities,
we have $t^*\ge 1$ and
\[
\sigma_i\left(a_i\cdot\frac{y^*}{t^*}+c_i\right)
\ge \frac{1}{t^*}>0
\qquad\text{for every }i.
\]
Thus, the generated point
$\theta=y^*/t^*$ belongs to the chosen region.
\end{proof}

Note that the generated parameters do not necessarily satisfy $w_j=\|z_j\|^2$ for every $j$. However, we use these parameters only to choose a minimum-score assignment in each block. After combining these assignments, we compute the actual cost of the resulting partition using the mean of each cluster.

By \cref{lem:local-choices,lem:comparison-samples},
at least one generated sample yields an optimal
combination. For each candidate, we split clusters if necessary to obtain exactly \(k\) nonempty clusters. We then compute its cost using the mean of each cluster and keep the partition with minimum cost.

\section{Analysis of the Algorithm}
\subsection{Correctness}
Let $\operatorname{OPT}$ be the minimum cost of a partition
of $P$ into $k$ nonempty clusters. We assume that
the input points are distinct and have integer coordinates in $\mathbb{R}^d$.

We first show that the algorithm always returns
a valid $k$-partition of $P$. Recall that the algorithm produces $t$ clusters on $Z$,
$t_B$ clusters on $B$, and $t_i$ small clusters formed
from the label-zero points in each block $X_i$.
It also produces at most $q$ additional clusters by
combining all points with the same positive label. Thus, every point belongs to exactly one cluster,
and the number of nonempty clusters is at most
\[
t+t_B+\sum_{i=1}^{f}t_i+q=k.
\]
If fewer than $k$ clusters are obtained, we can
repeatedly split a cluster containing more than one point
until there are exactly $k$ clusters, without increasing
the total cost. Therefore, every candidate gives a valid
$k$-partition whose cost is at least
$\operatorname{OPT}$.

We next show that at least one of the choices
enumerated by the algorithm preserves an optimal solution.

Fix an optimal partition of $P$.
By Section~\ref{sec:size-gap}, there exists a size
interval $(L,U]$ such that the union $Z$ of all
clusters whose sizes belong to this interval
contains at most $n/K$ points.
The algorithm enumerates this interval, the set $Z$,
and the number $t$ of clusters on $Z$.
After removing $Z$, every remaining cluster has
size at most $L$ or greater than $U$.

By Lemma~\ref{lem:crossing}, there exists a grid and a set $B$ such that $B$ is
a union of small clusters, and every
remaining small cluster lies entirely within
one block $X_i$.

Now let $q$ be the number of large clusters,
and let $t_B,t_1,\ldots,t_f$ be the numbers of small
clusters in $B$ and the blocks.
The algorithm enumerates these numbers.
We assign a distinct positive label to each
large cluster and label zero to all points
belonging to small clusters.
This gives a feasible assignment in every block.

By Lemma~\ref{lem:baseline}, the algorithm solves
$Z$, $B$, and the label-zero points in each block
optimally for their respective numbers of clusters.
Thus, their costs cannot exceed those in
the original optimal partition.
The points with the same positive label form the
corresponding large cluster in the original optimal
partition.
Therefore, there exists a combination of feasible
assignments such that its total cost, together with
the costs of the clusters on $Z$ and $B$, is at most
$\operatorname{OPT}$.

It remains to show that the algorithm can find
such a combination of assignments.

By Lemma~\ref{lem:local-choices}, there exist
parameters $z_1,\ldots,z_q\in\mathbb{R}^d$
and $w_1,\ldots,w_q\in\mathbb{R}$ such that choosing any
minimum-score assignment in each block gives
a combination with minimum total cost among
all feasible combinations. 

Let $\theta^*=(z_1,\ldots,z_q,w_1,\ldots,w_q)$
denote the vector containing the parameters guaranteed
by Lemma~\ref{lem:local-choices}.
Choose an open region whose closure contains
$\theta^*$. Within this region, the order of the scores
does not change. Since the scores are continuous, any assignment with minimum score in
this region also has minimum score at $\theta^*$.

By Lemma~\ref{lem:comparison-samples}, the algorithm
generates a parameter point in every nonempty open
region.
Within the same region, the order of the scores
does not change.
Thus, the algorithm considers a combination of
minimum-score assignments at the parameters
guaranteed by Lemma~\ref{lem:local-choices}.
This combination has minimum total cost among all
feasible combinations.

By the preceding argument, for the correct choices
of $Z$, $B$, the blocks, and the cluster counts,
the resulting total cost is at most $\operatorname{OPT}$.
Since every candidate gives a valid $k$-partition,
its cost cannot be smaller than $\operatorname{OPT}$.
Therefore, the algorithm returns a partition
with cost exactly $\operatorname{OPT}$.

\subsection{Running Time and Space}
\label{sec:running-time}

We first establish several bounds on the parameters.
\subparagraph{Bounds on the parameters.} Fix one of the $K$ size intervals $(L,U]$,
where every small cluster has size at most \(L\), and every large cluster has size greater than \(U\).
Recall that the grid construction uses
\[
a=\left\lfloor L^{1/[2(d+1)]}\right\rfloor,
\qquad
s=L^{\lceil(d+3)/2\rceil}.
\]
Here, $a$ controls the number of points removed
during the grid construction, while \(s\) is an upper bound on the size of each block.
Since
\[
\left\lceil\frac{d+3}{2}\right\rceil+1
\le d+2
\]
and $a\le L$, we have
\[
sa\le L^{d+2}\le U.
\]

Let $m=|P\setminus Z|\le n$.
Recall that every choice of $q$ considered by the
algorithm satisfies
\[
q(U+1)+(k-t-q)\le m.
\]
Since $q\le k-t$, we have $qU\le m$. Together with $sa\le U\le n$, this gives $s\le n/a$ and $qs\le n/a$.

Moreover, since $a\le s$ and the number of blocks
satisfies $f\le 2m/s+1$, we have
\[
f\le\frac{2n}{a}+1.
\]

 We first analyze the enumeration of the size
intervals and the set $Z$.
Recall that there are $K$ size intervals.
For each interval, the algorithm enumerates
all subsets $Z\subseteq P$ with
$|Z|\le\lfloor n/K\rfloor$.
Thus, the number of choices for $Z$ is at most
\[
\sum_{j=0}^{\lfloor n/K\rfloor}\binom{n}{j}.
\]

For each $Z$, the algorithm considers at most
$n+1$ choices for its number $t$ of clusters.
By Lemma~\ref{lem:baseline}, solving $Z$ with
$|Z|=j$ takes
\[
O^*\left(2^{3j+O(\log^2(n+2))}\right)
\]
time.

We next analyze the running time for the remaining
points after fixing the size interval $(L,U]$ and
the set $Z$.
Let $X=P\setminus Z$ and $m=|X|$.

\subparagraph{Enumerating the grids and cluster counts.} As described in Section~\ref{sec:small-clusters},
the number of possible cell side lengths is polynomial
in the input encoding length.
For a fixed cell side length, the assignment of points
to cells changes only when a grid boundary passes an input point.
Thus, each coordinate direction has at most $m+1$
relevant shifts, giving at most $(m+1)^d$ shifts
in total.

As shown in Section~\ref{sec:small-clusters}, for a suitable grid,
at most $2dm/a$ points belong to small clusters crossing cell
boundaries, and at most $\varepsilon_d m/a$ additional points
belong to small clusters contained in heavy cells, where $\varepsilon_d=1+\left\lceil(1+\sqrt{2d})^d\right\rceil$. For each grid, the algorithm enumerates all
subsets $B\subseteq X$ satisfying
\[
|B|\le b=\min\{m,\lfloor c_dn/a\rfloor\},
\]
where $c_d=2d+\varepsilon_d$.
The number of such subsets is at most
\[
\sum_{j=0}^{b}\binom{m}{j}
\le (m+1)^{b+1}.
\]

For each set $B$, the algorithm enumerates
the numbers $q,t_B,t_1,\ldots,t_f$.
Since each number is at most $n$, there are
at most
\[
(n+1)^{f+2}
\]
such choices.

For each choice, the algorithm solves the
instance on $B$ using Lemma~\ref{lem:baseline}.
Since $|B|\le b$, this takes
\[
O^*\left(2^{3b+O(\log^2(n+2))}\right)
\]
time.

Since $b\le \frac{c_dn}{a}$ and $f\le \frac{2n}{a}+1$,
the total time for enumerating the grids, choosing $B$,
enumerating the cluster counts, and solving $B$ is at most
\[
O^*\left((n+2)^{O(d+c_dn/a)}\right).
\]

Next we analyze the time of generating the assignments in each block and finding a minimum-score assignment for each block.

\subparagraph{Processing the blocks.} We assume that the grid, $B$, and the cluster counts are known.
Each block contains at most $s$ points, and each
point can be assigned one of the $q+1$ labels.
Thus, each block has at most $(q+1)^s$ assignments.
Since there are $f$ blocks, the total number of
assignments is at most
\[
M=f(q+1)^s.
\]

We first bound the number of parameter points generated by the algorithm.
Recall that $r=(d+1)q$ is the number of parameters.
Each pair of assignments in the same block defines
at most one hyperplane. Since there are at most $M$ assignments
in total, there are at most $\binom{M}{2}$ hyperplanes.

Each hyperplane gives two row vectors, and we add one additional
row. Thus, there are at most $M(M-1)+1$ rows.
The algorithm enumerates sets of at most $r+1$ rows and skips
those whose rows are linearly dependent.
Therefore, the number of sets considered is at most
\[
\sum_{j=0}^{r+1}\binom{M(M-1)+1}{j}
\le (M(M-1)+2)^{r+1}
\le (M+r+2)^{2r+2}.
\]

Next, we analyze the time of processing one
set of rows.

Except for the additional row, each row is obtained from two assignments.
To compute the coefficients of an assignment,
the algorithm solves the instance on 
points with zero label using Lemma~\ref{lem:baseline}.
Since each block contains at most $s$ points,
this takes
\[
O^*\left(2^{3s+O(\log^2(n+2))}\right)
\]
time per assignment.
The remaining coefficients can be computed
in polynomial time.

Since each set contains at most $r+1$ rows,
generating these rows requires computing the
coefficients of at most $2(r+1)$ assignments.

The algorithm then checks whether the selected
rows are linearly independent and solves the
corresponding linear system.
Since the system has at most $r+1$ equations,
checking linear independence and solving the
linear system take polynomial time.

After computing a parameter point, the algorithm checks whether
it lies on any of the hyperplanes. This check examines at most
$\binom{M}{2}$ pairs of assignments. If the point lies on a
hyperplane, the algorithm discards it. For each generated parameter point, the algorithm
examines at most $M$ assignments and chooses one
with minimum score in each block.
It then combines the selected assignments and
computes their total cost in polynomial time.

Therefore, processing one set of rows requires at most
$2(r+1)$ assignment evaluations to generate the rows,
$2\binom{M}{2}$ to check the hyperplanes, and $M$ to choose
a minimum-score assignment in each block.
Its running time is thus at most
\[
O^*\left(
(M^2+2r+2)\,2^{3s+O(\log^2(n+2))}
\right).
\]

Multiplying this bound by the number of row sets,
we obtain a total running time of
\[
O^*\left(
(M+r+2)^{2r+4}\,2^{3s+O(\log^2(n+2))}
\right).
\]

Recall that
\[
qs\le\frac{n}{a},
\qquad
s\le\frac{n}{a},
\qquad
f\le\frac{2n}{a}+1.
\]
Suppose that $q\ge1$. Since $d=(\log n)^{o(1)}$, we have
$d\le n$ for all sufficiently large $n$.
Using $M=f(q+1)^s$, $q\le n$, and $r=(d+1)q$, we obtain
\[
\log(M+r+2)=O\bigl(s\log(n+2)\bigr).
\]
Since $r+1=O((d+1)q)$ and $qs\le n/a$, it follows that
\[
(r+1)\log(M+r+2)
=O\left((d+1)\frac{n}{a}\log(n+2)\right).
\]
The term $s$ is covered by this bound since $s\le n/a$.
The term $\log^2(n+2)$ is also covered, since
$a\le s$ and $sa\le n$ imply $n/a\ge\sqrt n$.
When $q=0$, the blocks can be solved independently
within the same time bound.

Therefore, the total running time for processing all blocks
is at most
\[
\boxed{
O^*\left((n+2)^{O((d+1)n/a)}\right).
}
\]

\subparagraph{Running time for a fixed interval and set \(Z\).} We have shown that after fixing a size interval $(L,U]$ and
$Z$, the total time for solving the remaining instance is bounded by
\[
O^*\left((n+2)^{O(c_dn/a)}\right).
\]
Here we use $c_d\ge d+1$ and $n/a\ge 1$ to combine the
bounds for enumerating the grids and processing the blocks.

We next show that this bound is $O^*(2^{o(n)})$ for every
one of the $K$ size intervals.
Recall that the intervals are constructed by repeatedly setting
\[
L=\lfloor U^{1/(d+2)}\rfloor,
\]
starting with $U=n$.
Thus, in the $j$-th interval, where $1\le j\le K$, we have
\[
L=\left\lfloor n^{1/(d+2)^j}\right\rfloor.
\]
Since $a=\lfloor L^{1/[2(d+1)]}\rfloor$, it follows that
\[
a=
\left\lfloor
n^{1/[2(d+1)(d+2)^j]}
\right\rfloor.
\]

By the definition of $K$, we have
\[
(d+2)^j\le (d+2)^K\le \sqrt{\log_2 n}.
\]
Using $\lfloor x\rfloor\ge x/2$ for $x\ge1$, we obtain
\[
\log_2 a
\ge
\frac{\log_2 n}{2(d+1)(d+2)^j}-1
\ge
\frac{\sqrt{\log_2 n}}{2(d+1)}-1.
\]

Recall that $c_d=2d+1+\lceil(1+\sqrt{2d})^d\rceil$. Thus,
$\log_2 c_d=O(d\log_2(d+2))$.
Since $d=(\log n)^{o(1)}$, we have
\[
\log_2\!\bigl(c_d\log_2(n+2)\bigr)
=
O\!\left(d\log_2(d+2)+\log_2\log_2(n+2)\right)
=
o\!\left(\frac{\sqrt{\log_2 n}}{d+1}\right).
\]
Together with the lower bound on $\log_2 a$, this gives
\[
\frac{c_d\log_2(n+2)}{a}\longrightarrow 0.
\]
Therefore,
\[
\frac{c_dn}{a}\log_2(n+2)=o(n),
\]
and the running time for the remaining instance is
$O^*(2^{o(n)})$.

\subparagraph{Total running time.} We now include the time for enumerating and solving $Z$.
There are $K$ size intervals and at most $n+1$
choices for the number of clusters in $Z$.
For $|Z|=j$, Lemma~\ref{lem:baseline} gives a running time of
\[
O^*\left(2^{3j+O(\log^2(n+2))}\right).
\]
Since $K=O(\log\log n)$ and $\log^2(n+2)=o(n)$,
the total running time is at most
\[
O^*\left(
2^{o(n)}
\sum_{j=0}^{\lfloor n/K\rfloor}
\binom{n}{j}8^j
\right).
\]

Let $H(x)=-x\log_2 x-(1-x)\log_2(1-x)$ denote the binary
entropy function (e.g., see~\cite[\S\,3.2]{FominK10}). Since $K\ge2$, we have
\[
\begin{aligned}
\sum_{j=0}^{\lfloor n/K\rfloor}\binom{n}{j}8^j
&\le
2^{3n/K}
\sum_{j=0}^{\lfloor n/K\rfloor}\binom{n}{j}\\
&\le
2^{(H(1/K)+3/K)n}.
\end{aligned}
\]
Since $d=(\log n)^{o(1)}$, we have
$\log_2(d+2)=o(\log_2\log_2 n)$. Thus,
\[
\frac{\log_2\log_2 n}{2\log_2(d+2)}\longrightarrow\infty.
\]
Taking the floor gives $K\to\infty$.
It follows that $H(1/K)+3/K=o(1)$.
Therefore, the total running time is
\[
\boxed{O^*(2^{o(n)}).}
\]

\subparagraph{Space usage.} We enumerate the size intervals, grids, subsets, and cluster counts one choice at a time. To generate a parameter point, we store at most $r+1$ rows. Apart from the additional row, each row is specified by a block, two assignments in that block, and a sign. 

The resulting linear system has at most $r+1$ equations,
where $r=(d+1)q\le(d+1)n$.
We next show that its coefficients have polynomial bit length.
After scaling the input to integer coordinates, all coordinates
have polynomial bit length.
For each assignment, since $h_{ij}$ and $u_{ij}$ are counts and sums
of input coordinates, they have polynomial bit length.

By Equation~(1), the cost of any nonempty cluster $C$
has a denominator dividing $|C|$.
Thus, the denominator of $b_i$ divides $n!$.
Since $\log(n!)=O(n\log n)$, the coefficients of the
affine scores and the coefficients of the linear system
have polynomial bit length.

Since the system has polynomial size and its coefficients
have polynomial bit length, we can solve it using
polynomial space.

We consider one parameter point at a time.
For each block, we keep only an assignment with
the lowest score found so far.
We then combine the selected assignments and update
the best partition if the new one has lower cost.
All rows are generated when needed, and all calls
to Lemma~\ref{lem:baseline} are performed sequentially.
Therefore, the algorithm uses polynomial space.
\section{Conclusion}

In this paper, we gave a deterministic exact algorithm for continuous Euclidean $k$-means running in $O^*(2^{o(n)})$ time and polynomial space when $d=(\log n)^{o(1)}$. Our algorithm works for arbitrary $k$, even when $k$ grows linearly with $n$.

Our approach shows that small and large clusters can be handled in different ways. By creating a gap in cluster sizes, we can use a grid to ensure that the remaining small clusters lie entirely within individual blocks. Large clusters may span several blocks, but their assignments can be combined using a common set of parameters. This allows us to find an optimal solution without explicitly enumerating all combinations of assignments across the blocks.

Our algorithm allows arbitrary rational input coordinates given in binary, with polynomial dependence on their encoding length. However, we have not established the same running time for arbitrary real coordinates in the real-RAM model. Extending the result to this setting would be interesting.

Another natural direction is to extend our algorithm to arbitrary dimensions. In higher dimensions, both the geometric bounds and the number of parameters used to combine the blocks become more expensive. Finding a more efficient way to combine assignments across blocks could potentially lead to an exact algorithm with running time strictly below $2^n$ for arbitrary dimensions.

\subparagraph{Disclosure about use of AI.} The results of the paper were obtained with the assistance of GPT-6. The
core idea of the algorithm was generated by the AI tool. We complemented some details, simplified many parts, and presented the solutions in our own way.  The AI tool was also used to check numerical calculations and to improve the English presentation. All mathematical claims and proofs were independently verified by
the authors, who take full responsibility for the content of this paper.

\appendix
\section{Preprocessing Rational Coordinates}
\label{app:rational-inputs}

Let $\ell$ be the total binary input length, and let $D$ be
the product of the denominators of all input coordinates.
We multiply every coordinate by $D$ to obtain a set of
integer points. Since the binary length of $D$ is at most $\ell$,
this transformation takes polynomial time and space.
For any cluster $C$, its mean is also multiplied by $D$.
Thus, the squared distance from each point to its cluster
mean is multiplied by $D^2$. Hence, every partition cost is
multiplied by the same factor. Therefore, an optimal partition of
the scaled points gives an optimal partition of the original points.

We also verify the range of grid scales used in
Section~\ref{sec:small-clusters}.
After scaling, any two distinct input points are at distance
at least $1$. If a cluster has positive radius $R$, it contains
two distinct points. Their distance is at most $2R$. Thus,
$R\ge 1/2$. Let $W$ be the maximum absolute value of a scaled
coordinate. Since every cluster mean lies in $[-W,W]^d$,
$R\le 2W\sqrt d$. Since $W$ has polynomial bit length,
enumerating powers of two
$\rho=1,2,4,\ldots$ up to a value exceeding the
maximum possible radius
takes polynomial time.
One of these powers $\rho$ satisfies $R\le\rho\le 2R$.

\newpage

\bibliography{lipics-v2021-sample-article}
\end{document}